\documentclass[iop, numberedappendix]{emulateapj}
\usepackage{graphicx}  
\usepackage{dcolumn}   
\usepackage{bm}        
\usepackage{amssymb}   
\usepackage{color}

\def\apj{Astrophys. J.}
\def\apjl{Astrophys. J. Lett.}
\def\apjs{Astrophys. J. Supp. Ser. }

\def\mnras{Mon. Not. Roy. Astron. Soc. }

\def\prl{Phys. Rev. Lett.}
\def\prd{Phys. Rev. D.}

\def\be{\begin{equation}}
\def\ee{\end{equation}}
\def\ba{\begin{eqnarray}}
\def\ea{\end{eqnarray}}
\def\go{\mathrel{\raise.3ex\hbox{$>$}\mkern-14mu
             \lower0.6ex\hbox{$\sim$}}}
\def\lo{\mathrel{\raise.3ex\hbox{$<$}\mkern-14mu
             \lower0.6ex\hbox{$\sim$}}}
\def\be{\begin{equation}}
\def\ee{\end{equation}}
\def\bea{\begin{eqnarray}}
\def\eea{\end{eqnarray}}
\newcommand{\bes}{\begin{subequations}}
\newcommand{\ees}{\end{subequations}}

\begin{document}

\title[Shattering Flares]{Shattering Flares During Close Encounters of Neutron Stars}
\author{David Tsang}\email{dtsang@physics.mcgill.ca} \affiliation{Department of Physics, McGill University, Montreal, QC, Canada}

\date{\today}

\begin{abstract}
We demonstrate that resonant shattering flares can occur during close passages of neutron stars in eccentric or hyperbolic encounters. We provide updated estimates for the rate of close encounters of compact objects in dense stellar environments, which we find are substantially lower than given in previous works. While such occurrences are rare, we show that shattering flares can provide a strong electromagnetic counterpart to the gravitational wave bursts expected from such encounters, allowing triggered searches for these events to occur.
\end{abstract}


\maketitle

\section{Introduction}
The major expected source of gravitational waves for the Advanced LIGO \citep{LIGOref} class of gravitational wave detectors are compact binary systems. The long inspiral signals from such binaries will be detected by matched filtering with theoretical templates, which allows signal-to-noise to be built up over many orbits \citep{Cutler1994}.

Gravitational waves (GWs) are also emitted as broad band bursts when compact objects undergo close passages, either during single parabolic or hyperbolic encounters, or during repeated eccentric encounters \citep{Kocsis2006}. Such events occur rarely, but are more likely in dense stellar environments, such as globular clusters or galactic nuclear clusters. The brief duration of such bursts do not allow a large integrated buildup of signal to noise, and they may be difficult to detect without some electromagnetic trigger.

Recently, \citet{Tsang2012} showed that during binary inspiral of neutron stars (NSs) resonant tidal excitation of the interface mode -- a natural mode of a neutron star peaked at the crust-core boundary -- could result in an isotropic resonant shattering flare, and that these were consistent with short Gamma-Ray Burst (sGRB) precursors observed seconds before some sGRBs \citep{Troja2010}. Coincident timing of such precursor flares and the GW inspiral signal can be used to provide strong constraints on the NS equation of state \citep{Tsang2012}. 

In this paper we show that resonant shattering flares can also occur during close passages with other compact objects, such as another NS or a black hole (BH), and that such flares could serve as electromagnetic counterparts to gravitational wave bursts, allowing triggered searches for these bursts.

\section{Tidal Energy Transfer During Parabolic and Eccentric Encounters}
Tidal energy transfer during close encounters can be determined in a Newtonian approximation through the procedure  outlined in \citet{Press1977a}. While a fully relativistic formulation would be preferable, the Newtonian formulation is sufficiently accurate for periapse distance much larger than the neutron star radius, relativistic effects would only slightly modify the frequencies and increase the strength of the interaction.

In general the energy transfer rate to a star is given by
\be
\frac{dE}{dt} = \int d^3 x \rho {\bm v} \cdot {\bm \nabla} U
\ee
where the fluid velocity ${\bm v} \equiv \partial{\bm \xi}/\partial t$ is the time derivative of the Lagrangian displacement ${\bm \xi}$ and $U$ is the gravitational potential. To examine the response of a neutron star with mass $M_1$ it is convenient to decompose the potential due to a star with mass $M_2$ into spherical harmonics ${Y}_{lm} (\theta, \phi)$,
\ba
U(r, \theta, \phi) = \sum_{l=0}^{\infty} \sum_{m=-l}^{l} U_{lm} {Y}_{lm}{}^*(\theta, \phi)\\
U_{lm} = W_{lm} \frac{GM_2 r^l}{R(t)^{l+1}} e^{im\Phi(t)}
\ea
where $\Phi$ is the true anomaly of the system, ${}^*$ denotes the complex conjugate, $R(t)$ is the distance between the stars, and $(r, \theta, \phi)$ is the co-moving coordinate system centered with $M_1$. Note that for the purposes of mode excitation we are only concerned with the tidal ($l > 2$) component of the potential. Assuming the normalization for the spherical harmonics given in \citet{Jackson}, the constants $W_{lm}$ are 
\ba
W_{lm} &=& (-)^{(l+m)/2} \left[\frac{4\pi}{2l+1}(l-m)!(l+m)! \right]^{1/2} \nonumber \\
&\quad& \times \left[ 2^l \left(\frac{l-m}{2}\right)!\left(\frac{l+m}{2} \right)!\right]^{-1}
\ea
where $(-)^k$ is defined to be zero when $k$ is a non-integer.

The energy transferred to a particular mode (assuming no non-linear effects, such as crust fracture) during a periapse passage for a parabolic or highly eccentric encounter can be estimated by eq (40) of \citet{Press1977a}
\be
\Delta E_{nlm} = 2\pi^2 \frac{GM_1^2}{R_1} \left(\frac{M_2}{M_1}\right)^2 \left(\frac{R_1}{R_{\rm min}} \right)^{2l + 2} |Q_{nl}|^2 |K_{nlm}|^2 
\ee
where $Q_{nl}$ is the overlap integral for the NS displacement eigenmode ${\bm \xi}_{nlm}$, with radial mode number $n$, 
\be
Q_{nl} \equiv\frac{1}{M R^2}\int d^3 x \rho~ \boldsymbol{\xi}_{nlm}{}^* \cdot 
\boldsymbol{\nabla}[ r^2 {Y}_{l,m}(\theta, \phi)],
\label{eq:q}
\ee
and 
\be
K_{nlm} = \frac{W_{lm}}{2\pi} 2^{3/2} \hat{\eta} I_{lm}(\hat{\omega}_{nlm})
\ee
\ba
I_{lm}(\hat{\omega}_{nlm}) \equiv \int_0^\infty (1 + x^2)^{-l}\cos[2^{1/2} \hat{\omega}_{nlm} (x + x^3/3)\nonumber\\
\quad ~ + 2 m \tan^{-1} x]dx.
\ea
Here $\hat{\eta} \equiv [M_1/(M_1 + M_2)]^{1/2} (R_{\rm min}/R_1)^{3/2}$ and $\hat{\omega}_{nlm} \equiv \omega_{nlm}  (M_1 + M_2)^{-1/2}R_{\rm min}^{3/2}$ are the Keplerian frequency at the neutron star surface and the mode frequency respectively, both scaled by the Keplerian frequency at periapse.

Modes with frequency much higher than periapse Keplerian frequency ($\hat{\omega}_{nlm} \gg 1$) cannot be strongly excited. In contrast if the the periapse distance is too small, the stars may collide, or a tidal disruption may occur. 

We can calculate the energy transfer to the interface mode $\Delta E_{i}$, by utilizing the i-mode frequencies and overlap integrals for various equations of state given by \citet{Tsang2012} and summarized in Table 1. If $\Delta E_{i} > E_b$, the mode energy required for the crust to reach the breaking strain, then a shattering flare can occur. $\Delta E_i/E_b$ for various equations of state are shown as a function of periapse distance $R_{\rm min}$ for parabolic encounters between $1.4 M_{\odot}$ neutron star  (Figure 1) and between a $10 M_\odot$ black hole and a $1.4 M_\odot$ neutron star (Figure 2). 

\begin{figure}
\includegraphics[width=\columnwidth]{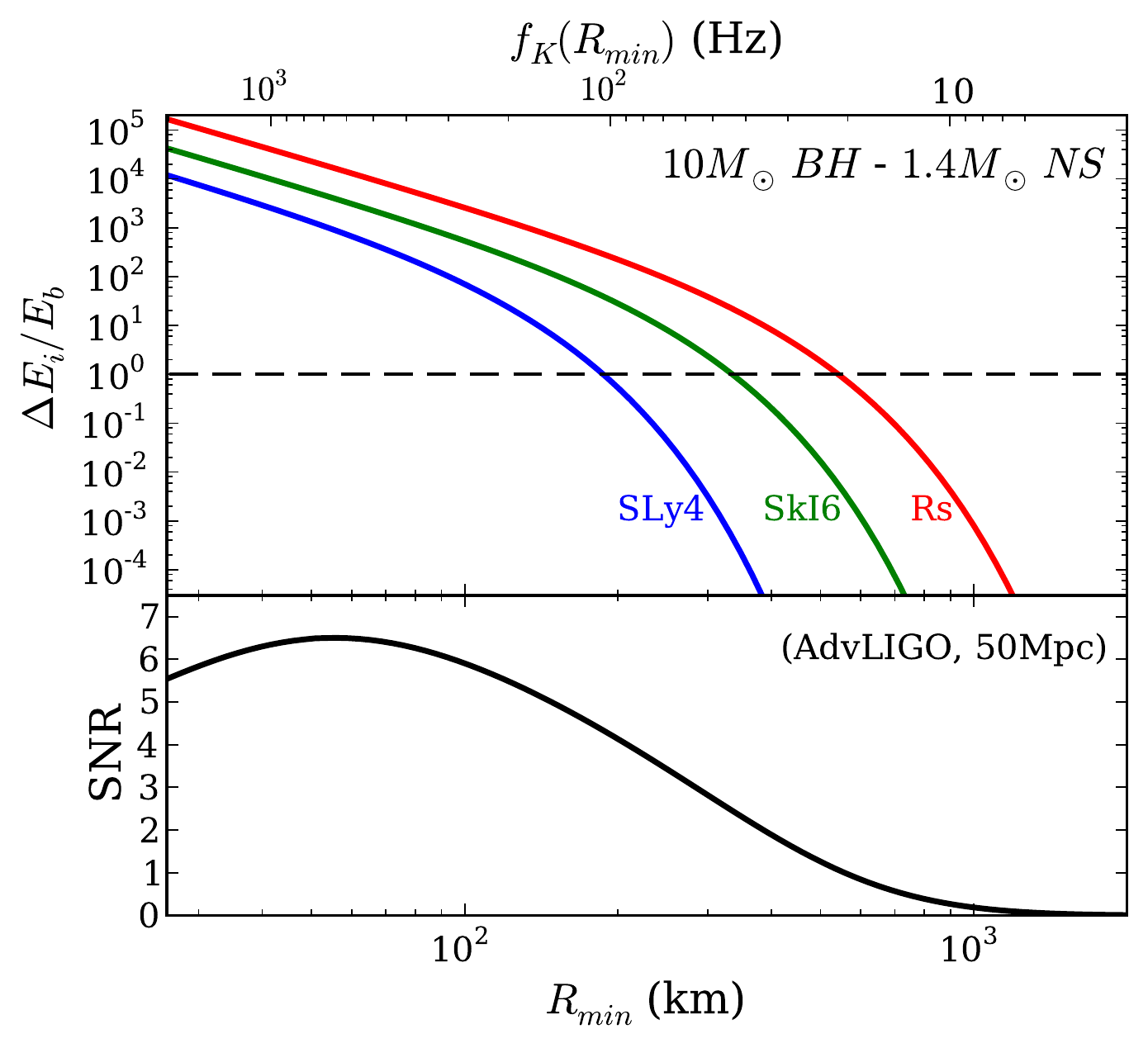}
\caption{Above: The ratio of the maximum energy transfer through tidal resonance to the interface mode, $\Delta E_i$, to the mode energy required to reach the breaking strain, $E_b$,  for a parabolic or highly eccentric $10 M_\odot$ BH- $1.4 M_\odot$ NS  encounter as a function of periapse distance ($R_{\rm min}$), for various equations of state. Below a critical periapse distance the mode energy exceeds the breaking energy  ($\Delta E_i/E_b > 1$) and a resonant shattering flare could occur. The upper axis shows the Keplerian orbital frequency at the periapse, $f_K(R_{\rm min}) \equiv [G(M_1 + M_2)/R_{\rm min}^3]^{1/2}/2\pi$. Below: The sky and observer inclination averaged signal to noise ratio (SNR) for a single Advanced LIGO gravitational wave detector for a gravitational wave burst at 50 Mpc from a parabolic 10 $M_\odot$ - 1.4 $M_\odot$ close encounter with periapse $R_{\rm min}$.}
\end{figure}

\begin{table}
\begin{center}
\begin{tabular}{ |l| c| c| c| r |} \hline 
EOS  & $R_{1.4}$ [km] &$f_{i}$ [Hz] &$Q_i$ &  $E_b$ [erg] \\ \hline
SLy4 & $11.7$ &$188$ & $0.041$ & $5\times 10^{46}$   \\
 SkI6 &$12.5$&$ 67.3$ & $0.017$& $3\times 10^{45}$  \\
 Rs & $13.0$ &$32.0$ & $0.059$ & $1\times 10^{46}$  \\
 \hline
\end{tabular}
\end{center}
\caption{Resonant mode properties for the $l=2$ i-mode for various equations of state from \citet{Tsang2012}. $R_{1.4}$ is the radius of a $1.4 M_\odot$ NS, $f_{i}$ is the i-mode frequency, $Q_i$ is the overlap integral for the i-mode and the tidal field, while $E_b$ is the mode energy required to reach the breaking strain in the crust.}
\label{tab:1}
\end{table}

\section{Resonant Shattering}

The process that produces a resonant shattering flare is outlined in Figure \ref{Fig:Flow}. During a close encounter (or at orbital resonance) energy is extracted from the kinetic energy of the orbit, through resonant tidal coupling. The interface mode is excited strongly, which drives the mode to an amplitude at which the breaking strain of the crust is exceeded. 

The crust fractures, depositing $\sim \epsilon_b \mu \Delta r^3 \sim 10^{43}$ erg of seismic energy into the crust, where $\epsilon_b \sim 0.1$ is the breaking strain \citep{Horowitz2009b}, $\mu$ is the shear modulus, and $\Delta r \sim 10^5$cm is roughly the thickness of the crust. These broad spectrum seismic waves are peaked at characteristic frequency $\sim (\mu/\rho)^{1/2}/(2\pi \Delta r) \sim 200$Hz, where $\rho$ is the density of the crust. Low frequency seismic waves cannot couple efficiently to the magnetic field \citep{Blaes1989}, and the energy builds up in the neutron star crust as the interface mode is driven further, and more fractures occur. This seismic energy builds until the crust reaches the elastic limit $E_{\rm elastic} = \int dV \epsilon_{b}^2 \sim 10^{46}$erg, when it shatters, scattering the mode and seismic energy to high-frequency oscillations which can then couple to the magnetic field. Strong perturbations of the magnetic field result in strong transverse electric fields, which can accelerate particles to high energy, sparking a pair-photon fireball. The luminosities of resonant shattering flares are expected to be up to $\sim 10^{47}-10^{48}$ erg/s \citep{Tsang2012}, if the precursor flare timescales are assumed.

\citet{Troja2010} found precursors occurring in 3 out of 49 SGRs analyzed, implying that not every binary merger should result in a detectable shattering flare.  We note that the extraction of seismic energy from the crust by the magnetic field is limited by the strength of the magnetic field at the surface of the neutron star. The maximum luminosity that can be extracted from the crust by the magnetic field can be estimated by
\be
L_{\rm max} \sim 10^{47} {\rm erg ~s}^{-1} (v/c) (B_{\rm surf}/10^{13}{\rm G})^2 (R/10^{10} {\rm cm})^2,
\ee
where $v$ is the maximum velocity of the perturbation to the field line, $R$ is the neutron star radius and $B_{\rm surf}$ is the local surface field strength, which can be significantly higher than the large scale dipole field. Thus only shattering flares from neutron stars with sufficiently strong surface fields can be detected. 

\begin{figure}
\includegraphics[width=\columnwidth]{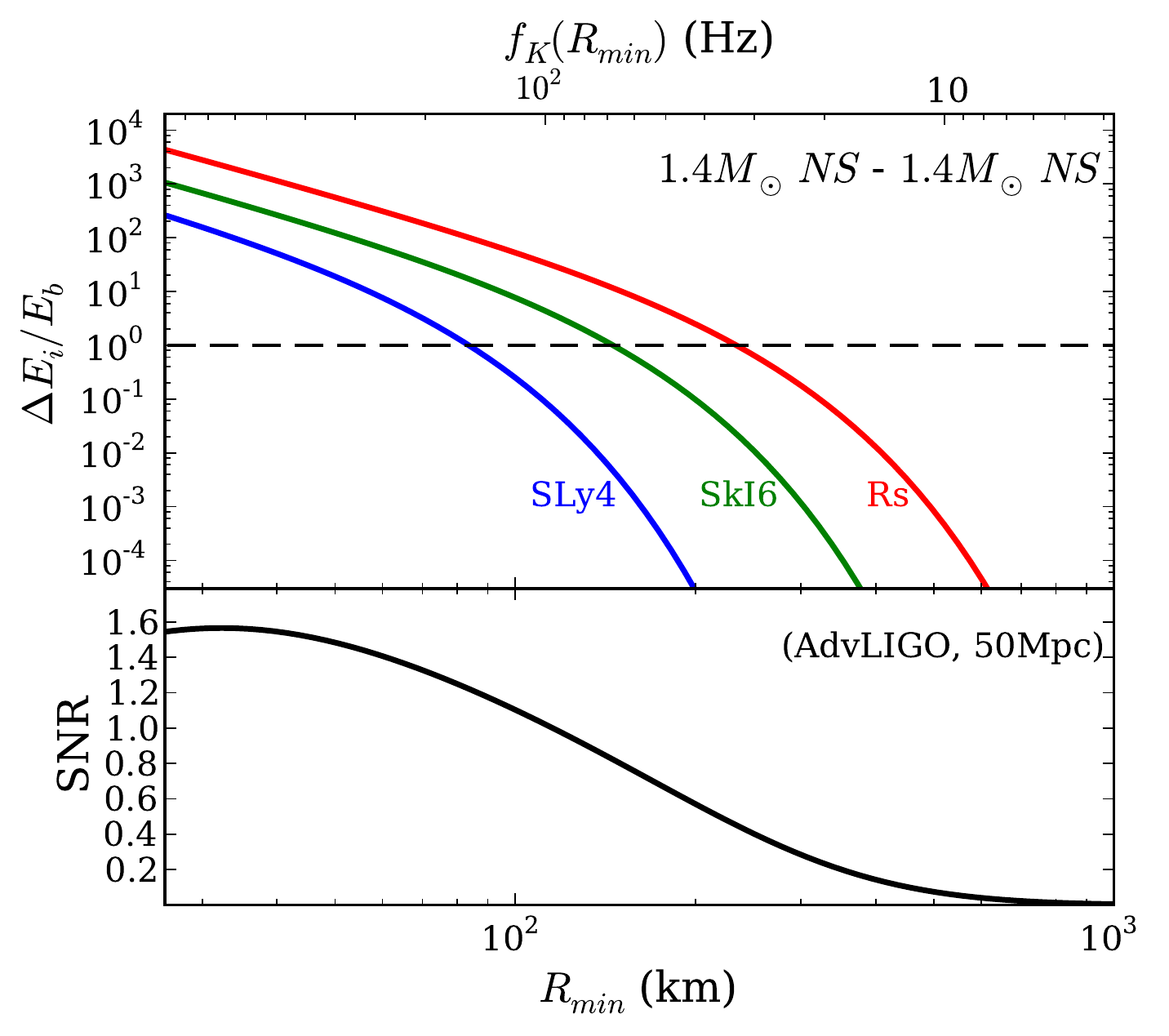}
\caption{Above: The ratio of the maximum energy transfer through tidal resonance to the interface mode, $\Delta E_i$, to the mode energy required to reach the breaking strain, $E_b$,  for a parabolic or highly eccentric $1.4 M_\odot$ NS- $1.4 M_\odot$ NS  encounter as a function of periapse distance ($R_{\rm min}$), for various equations of state. Below a critical periapse distance the mode energy exceeds the breaking energy  ($\Delta E_i/E_b > 1$) and a resonant shattering flare could occur. The upper axis shows the Keplerian orbital frequency at the periapse, $ f_K(R_{\rm min}) \equiv [G(M_1 + M_2)/R_{\rm min}^3]^{1/2}/2\pi$. Below: The sky and observer inclination averaged signal to noise ratio (SNR) for a single Advanced LIGO gravitational wave detector for a gravitational wave burst at 50 Mpc from a parabolic 1.4 $M_\odot$ - 1.4 $M_\odot$ close encounter with periapse $R_{\rm min}$.}
\end{figure}

\section{Electromagnetic Counterparts to Gravitational Wave Bursts}

To calculate the expected gravitational wave signal to noise ratio due to parabolic encounters we follow the procedure outlined in \citet{Kocsis2006}.
The strain due to a gravitational wave burst due to a parabolic encounter is given by \citep{Flanagan1998}
\be
h(f) = \frac{\sqrt{3}}{2\pi} \frac{G^{1/2}}{c^{3/2}} \frac{1+z}{d_L}\frac{1}{f} \sqrt{\frac{dE}{df}[(1 + z) f]}
\ee
where $z$ is the redshift, $d_L$ is the luminosity distance, and $dE/df$ is the total gravitational wave energy emitted by encounter per unit frequency, which is given for a parabolic ($e=1$) encounter in the non-relativistic limit by equation (46) from \citet{Turner1977}. The signal to noise ratio (SNR) for a sky and orientation averaged signal on a single detector is given by \citep{Dalal2006, Nissanke2010}
\be
{\rm SNR} = \frac{8}{5}\sqrt{ \int_0^\infty \frac{|h(f)|^2}{S_n(f)^2}df}
\ee
where $S_n(f)$ is the spectral noise density for a given detector. In Figures 1 and 2 the SNR for is shown for the NS-NS and BH-NS encounters assuming a single encounter at 50 Mpc ($z \simeq 0.011$), for advanced LIGO, with spectral noise density given by \citet{LIGOref}.  

Blind detection \citep[SNR $\go 6$ coincident at each detector, see e.g.][]{Aasi2013} of a single gravitational wave burst from a neutron star close encounter would be extremely challenging at reasonable distances, with fairly low SNR even for close passages, in particular for NS-NS encounters. Using X-ray or gamma-ray detections of resonant shattering flares as electromagnetic counterparts, triggered GW searches could be performed, significantly lowering the SNR threshold for GW burst detection \citep{Kochanek1993, Nissanke2010, Kelley2013, Dietz2013}. Networks of detectors can also be used to enhance burst detection, through coincident and coherent methods \citep{ Schutz2011,Nissanke2013, Aasi2013}.

\citet{Kocsis2012} also show that repeated GW bursts from eccentric captures can be combined with the final chirp to boost the integrated  SNR by roughly an order of magnitude, and would optimistically allow detection of bursts from BH-NS eccentric captures out to $\sim 300$ Mpc, and NS-NS encounters to $\sim 150$ Mpc. The pattern of these repeated bursts can be modelled for given orbital parameters. Resonant shattering flares can be seen significantly father than the GW bursts.  If they occur for a given system, they will happen for sufficiently close passages which are also those that contribute the largest component of the GW burst signal. If repeated flares are seen, these could also be used to characterize the orbit and target a burst search to accumulate SNR over multiple passages. However, significant changes to the current gravitational wave templates may be necessary to detect eccentric captures and mergers \citep{East2013, Huerta2013}.


\section{Event Rates}
Close encounters of neutron stars with other compact objects are much more likely to occur in dense stellar environments, such as globular clusters and galactic nuclei. While it is beyond the scope of this paper to perform an extremely detailed evaluation of the event rates for close encounters of compact objects, we will briefly discuss the event rates for such encounters in both these environments, and provide updated estimates for some of the rates in the literature. 

\subsection{Globular Clusters}
\citet{Kocsis2006} calculated the parabolic encounter rate for compact objects in globular clusters using simplified globular cluster models, predicting a rate of $\go 1$ detection per year for advanced LIGO in optimistic scenarios. However their detection rates are dominated by rare distant events involving close encounters of $\go 20 M_\odot$ black holes. 

\citet{Lee2010} examine various dynamical pathways to sGRBs in globular clusters, including binary interactions and tidal capture. They calculate a high rate of close encounters for two neutron stars, $\Gamma_{\rm NS-NS}^{(GC)} \sim 55$ yr$^{-1}$ Gpc$^{-3}$ using as a calibration the estimate of $\sim 10^4$ NSs in the collapsed core of M15, from the Fokker-Planck calculations of \citet{Dull1997}. This would require an extremely high neutron star retention fraction. Subsequent more careful calculations by \citet{Murphy2011} have determined the number of neutron stars in the core of M15 to be closer to $\sim 10^3$, consistent with $\sim 1-10\%$ retention fraction estimates from pulsar kick velocities \citep{Drukier1996, Hansen1997, Davies1998}. This reduces the estimates of \citet{Lee2010} to $\Gamma_{\rm NS-NS}^{(GC)} \sim 0.5$ yr$^{-1}$ Gpc$^{-3}$. 

\begin{figure}
\includegraphics[width=\columnwidth]{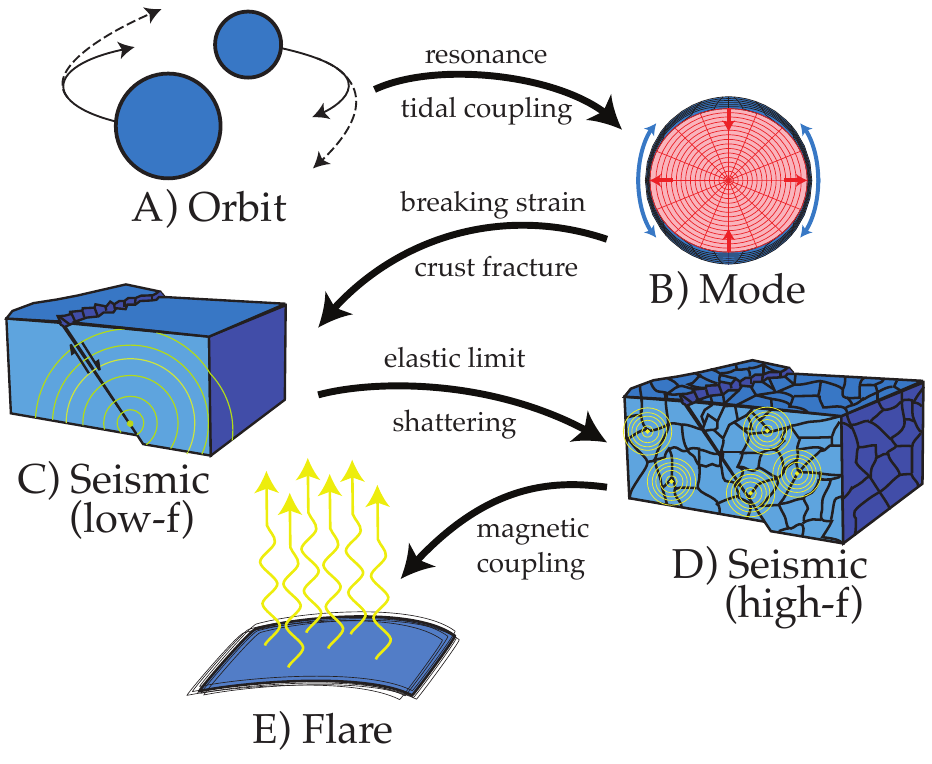}
\caption{A cartoon of the resonant shattering process. The gravitational potential of the system is the ultimate source of the energy powering the resonant flare. During close passage, or at resonance for circular orbits, tidal resonance transfers energy from the orbit (A) to the i-mode (B) at a rate $ \go 10^{50}$ erg/s. The i-mode grows quickly until the maximum strain at the base of the crust exceeds the breaking strain at mode energy $\sim 10^{47}$ erg. A fracture occurs, releasing $\sim 10^{43}$ erg of low frequency seismic energy (C) per fracture, however the mode continues to be driven by the resonance. As more fractures occur, more energy is deposited into seismic energy in the crust. When the total seismic energy in the NS crust exceeds the elastic limit of the curst $E_{\rm elastic}  \sim 10^{46}$ erg, the crust shatters, scattering the mode energy and elastic energy to high frequency oscillations (D). High frequency oscillations can couple strongly to the magnetic field \citep{Blaes1989, Thompson1998} by strongly vibrating their footprints (E). Strong perturbations of the magnetic field at the neutron star surface drive strong electric fields, which can accelerate charged particles, triggering pair production and a relativistic fireball with luminosity $10^{47} - 10^{48}$ erg/s. \label{Fig:Flow}} 
\end{figure}

\subsection{Galactic Nuclei}
While there are many globular clusters per galaxy, high kick velocities at NS birth significantly lower their retention fraction. The deeper potentials of galactic nuclei may provide  dense stellar environments where the NS retention fraction is higher and close encounters are more likely to occur. \citet{OLeary2009}
and \citet{Kocsis2012} both provide estimates of $\sim 1-100$ black hole close passages per year within a few Gpc detectable by advanced LIGO, with GW-detectable NS-BH encounters estimated to be $\sim 1\%$ of this. However, as we discuss in the Appendix below, they scale by a factor $\xi$, representing the contribution due to the variance of the nuclear cluster density across galaxies. They take this factor to be  $\xi \go 30-100$, but we find $\xi$ is more correctly evaluated to be significantly lower, even with the most optimistic assumptions. 

In Appendix A we have re-evaluated the rates for single-single eccentric captures of compact objects in nuclear star clusters containing massive central black holes, assuming a simplified isothermal density distribution, as in \citet{Kocsis2012}. We find that the $10 M_\odot$ BH-BH eccentric capture encounter rate is $\Gamma_{\rm BH-BH}^{(GN, EC)}\sim 0.02 $ yr$^{-1}$ Gpc$^{-3}$, significantly lower than the previously estimated values. However a more top-heavy IMF \citep{Bartko2010} along with enhanced segregation and spatial flattening for heavier BHs may help to increase this value. 

The NS-NS and NS-BH (10 $M_\odot$) rates can be estimated in a similar fashion for an isothermal $\propto r^{-2}$ density distribution to be $\Gamma_{\rm NS-NS}^{(GN,EC)} \sim 0.04-6$ yr$^{-1}$ Gpc$^{-3}$, and $\Gamma_{\rm NS-BH}^{(GN,EC)} \sim 0.05-0.6$ yr$^{-1}$ Gpc$^{-3}$ with the range mainly due to uncertainty in the IMF and mass loss models for NS progenitors \citep{O'Connor2011}. However, there is reason to suspect that the slope of the density distribution is somewhat flattened due to interaction with segregated BHs \citep[see e.g.][and references therein]{OLeary2009}. Taking a neutron star density distribution $\propto r^{-3/2}$ as a lower bound for our rates we find in this case $\Gamma^{(GN,EC)}_{\rm NS-NS} \simeq 0.003-0.3$ yr$^{-1}$ Gpc$^{-3}$.

Note that in evaluating the above rates, we have taken very optimistic assumptions about systematic vs intrinsic variation of the relevant observations, as in \citet{OLeary2009}. Possibly more realistic estimates for this intrinsic scatter reduce these rates by a factor of $\sim 4$. 

In the high density, high relative-velocity region near the center of nuclear star clusters, there may be a significant rate of hyperbolic passages where the periapse is sufficiently close to trigger a shattering flare, but insufficient to result in eccentric capture by GW emission. In Appendix B we have calculated the rate of encounters that result in a shattering flare during the first close passage, and find this to be higher than the eccentric capture rate for the fiducial model used. We find, for the optimistic assumptions about intrinsic variation used above, and assuming our canonical isothermal model $\Gamma^{(GN, SF)}_{\rm NS-NS} \simeq 0.2 - 60$ yr$^{-1}$ Gpc$^{-3}$. For a more flattened density profile $\propto r^{-3/2}$ we have $\Gamma^{(GN, SF)}_{\rm NS-NS} \simeq 0.005 - 0.5$ yr$^{-1}$.
Taking our less generous estimates for the intrinsic scatter across galaxies significantly reduces these rates by a factor of $\sim 6$.

\subsection{Other Possible Event Rate Contributions}

In the above discussion we have primarily considered single-single interactions in determining the event rates in dense clusters. Binary-single or binary-binary interactions have larger cross sections and could increase the rates for such events \citep[see e.g.][]{Miller2009}. Recently \citet{Katz2012} and \cite{Kushnir2013} demonstrated that Kozai-Lidov type interactions can drive the inner binaries of hierarchical triple systems towards extreme eccentricity, with collisions occurring when the periapse distance is driven below the stellar radius. They claim that the rates for such Kozai-oscillation driven collisions between white dwarfs in field triple star systems can be comparable to the SN Type Ia rate. Similar interactions could potentially drive close encounters of NSs or BHs in triple systems, However, the periapse for shattering flares is two orders of magnitude smaller than those considered by \citet{Katz2012}, and NSs and BHs are substantially more rare than white dwarfs, particularly outside of dense clusters.  Within clusters such triple systems would need to reach these extreme eccentricities quickly, before other encounters ionize away the softer less-bound outer companion.

\section{Discussion}

We have calculated the energy transfer to the interface mode through tidal interaction for neutron stars during close encounters with other compact objects, and shown that resonant shattering flares can occur during parabolic or eccentric encounters if the periapse is sufficiently close and the local surface field of the NS is sufficiently high. Such flares are similar to resonant shattering flares during binary inspirals, and should have luminosity $\sim 10^{47}-10^{48}$ erg s$^{-1}$ \citep{Tsang2012}. 

Broad band gravitational wave bursts are also generated by such encounters. While they are rare, there is intense interest in such gravitational wave burst events which are detectable by the next generation of gravitational wave detectors. GW bursts with high SNR are also those for which shattering flares may occur and act as an electromagnetic counterpart to trigger burst searches. Highly eccentric captures of NSs in dense stellar environments are expected to result in repeated GW bursts \citep{Kocsis2012}, but may also lead to repeated shattering flares at each periapse passage. 

We have also reviewed and updated the rates for compact object encounters presented in the literature for dense stellar environments, and find that these rates should be revised significantly downwards.  While these estimated rates for close encounters involving neutron stars within the horizon of advanced LIGO are low, shattering flares during from such encounters can be detected significantly farther, and may occur at a rates of up to $\Gamma^{(GN, SF)}_{\rm NS-NS} \simeq 0.2 - 60$ yr$^{-1}$ Gpc$^{-3}$, subject to large model uncertainties. More conservative assumptions substantially lower this rate.

\section*{Acknowledgments}
This research was supported by generous funding from the Lorne Trottier Chair in Astrophysics and Cosmology as well as the Canadian Institute for Advanced Research, and brought to you by the letter $\xi$. I would like to thank  Bence Kocsis, Ryan O'Leary, and Cole Miller for detailed and enlightening discussions involving the rates, as well as Samaya Nissanke, Tony Piro, Enrico Ramirez-Ruiz, Kostas Gourgouliatos, Andrew Cumming, Sterl Phinney, and Gil Holder for insightful conversation. 

\appendix
\section{Eccentric Capture Rates in Nuclear Star Clusters}
Here we will carefully estimate the eccentric capture rate of compact objects in nuclear star clusters. We begin by following the general procedure outlined in the Appendix C of \citet{Kocsis2012}, and calculate the eccentric capture rate for a single fiducial galaxy.

\subsection{The Rate for a Single Galaxy}
The cross section for eccentric capture is basically the cross section for which the energy emitted by GW (or lost due to tidal interactions) during a close encounter exceeds the kinetic energy of the objects at infinity. The maximum periapse for capture is given by \citet{Quinlan1989} as
\be
r_{p, max} = \left[\frac{85 \pi \sqrt{2} G^{7/2} m_i m_j (m_i + m_j)^{3/2}}{12 c^5 |{\bm v}_i - {\bm v}_j|^2} \right]^{2/7} = 190 {\rm km} \left(\frac{\eta}{0.25} \right)^{2/7} \left(\frac{m_{tot}}{2.8 M_\odot} \right) \left(\frac{v_{\rm rel}}{10^3 {\rm km\, s}^{-1}}\right)^{-4/7}
\ee
where $\eta$ is the symmetric mass ratio, $m_{tot}$ is the total mass of the two objects, and $v_{\rm rel}$ is the relative speed of the two objects at infinity. We only consider gravitational wave capture here since, for compact objects, the tidal capture cross section is much smaller than the gravitational wave capture cross section.

We can then calculate the cross section, using the standard formula for gravitational focussing
\ba
\sigma_{cs} &=& \pi r_{p, max}^2 \left[1 + \frac{2Gm_{tot}}{r_{p, max} v_{\rm rel}^2} \right] \simeq \frac{2\pi G m_{tot} r_{p,max}}{v_{\rm rel}^2} \nonumber \\
 &=& 1.3 \times 10^{23} {\rm cm}^2 (\eta/0.25)^{2/7} (m_{\rm tot}/20 M_\odot)^2 (v_{\rm rel}/84 {\rm km s}^{-1})^{-18/7}
\ea
where we've scaled this to fiducial values that will be used later. The rate of eccentric captures for a single galaxy with central supermassive black hole of mass $M_S$ and velocity dispersion $\sigma_{\rm disp}(M_S)$ is then
\be
\Gamma_{\rm gal}(M_S) \simeq \int_{r_{\rm min}}^{r_i} dr 4\pi r^2 n_1(r, M_S) n_2(r, M_S) \sigma_{cs} v_{\rm rel}
\ee
where $n_{1}$ and $n_{2}$ are the number densities of each type of object as a function of radius, $r_i \equiv GM_S/\sigma_{\rm disp}^2$ is the radius of influence of the black hole and $r_{\rm min}$ is approximately the radius inside which there is only a single object.  

We take the density of objects to be the same as in an isothermal distribution, and the velocity distribution to be Maxwellian at each radius with relative speed $\langle v_{\rm rel}^2\rangle  = 2 v_{\rm circ}^2 \equiv 2GM_S/r = 2\sigma_{\rm disp}^2 (r/r_i)^{-1}$, which captures the behavior well near the central black hole, where the contribution to the rate is the largest. The number densities are then given by
\be
n_i(r) = \frac{N_i}{4 \pi r_{\rm dyn}^3}\left(\frac{r}{r_{\rm dyn}}\right)^{-2}
\ee
where $r_{\rm dyn}$ defines the dynamical radius inside which twice the mass of the central black hole is contained such that 
$N_{i} \equiv 2\kappa_i M_S/m_i$ and $\kappa_i$ are the number of objects and total mass fraction of type $i=1,2$ within $r_{\rm dyn}$ respectively. We take these scalings of $n(r)$ and $v_{\rm rel}(r)$ for simplicity to highlight the sources of uncertainty and provide a rate estimate for the simplest case. For how different and more realistic scalings may alter the basic single galaxy rate, see detailed discussion in \citet{OLeary2009} and Appendix C of \citet{Kocsis2012}.

It is also convenient to define the geometric means of the number $\tilde{N}$ and mass fraction $\tilde{\kappa}$, such that 
\be
\tilde{N}^2 \equiv N_1 N_2 =  \frac{4 \tilde{\kappa}^2 M_S^2}{\eta \,m_{\rm tot}^2}, \qquad \tilde{\kappa}^2 \equiv \kappa_1 \kappa_2,
\ee
as well as  the fiducial scalings $\sigma_{84} \equiv \sigma_{\rm disp}/(84 $km s$^{-1}$), $\eta_{0.25} \equiv \eta/0.25$, and $m_{20} \equiv m_{\rm tot}/(20 M_\odot)$. Approximating $r_{\rm min} \simeq \tilde{N}^{-1} r_{\rm dyn}$ we find
\ba
\Gamma_{\rm gal}(M_S) &\simeq& \int_{r_{\rm min}}^{r_i} dr 4\pi r^2 \frac{\tilde{N}^2}{(4\pi r_{\rm dyn}^3)^2}\left(\frac{r}{r_{\rm dyn}}\right)^{-4} \sigma_{cs} v_{\rm rel}\\
&\simeq& 1.3 \times10^{30}   {\rm cm}^3 {\rm s}^{-1} \eta_{0.25}^{2/7} m_{20}^2 \sigma_{84}^{-11/7} \left( \frac{r_i}{r_{\rm dyn}}\right)^{-11/14}   \frac{\tilde{N}^2}{4\pi r_{\rm dyn}^3}  \int_{\tilde{N}^{-1}}^{r_i/r_{\rm dyn}} dx  ~x^{-2 + 11/14}
\ea
where we have averaged over the Maxwellian distribution, $\langle v_{\rm rel}(r_i)^{-11/7} \rangle = 1.15 \times \sigma_{\rm disp}^{-11/7}$, and changed integration variables to $x \equiv r/r_{\rm dyn}$. Assuming that $(\tilde{N}r_i/r_{\rm dyn})^{3/14} \gg 1$, we have
\be
\Gamma_{\rm gal}(M_S) = 1.2\times10^{-10} {\rm yr}^{-1} (r_i/r_{\rm dyn})^{31/14}  \eta_{0.25}^{-23/28} \tilde{\kappa}_{2.5}^{31/14} m_{20}^{-3/14} M_{4e6}^{-11/14} \sigma_{84}^{31/7} \label{1galrate}
\ee
where $M_{4e6} \equiv M_S/(4 \times 10^{6} M_\odot)$ is scaled to the Milky Way, and $\tilde{\kappa}_{2.5} \equiv \tilde{\kappa}/(2.5\%)$ as in \citet{Kocsis2012}. 
Applying the $M-\sigma$ relation, $M_{4e6} =\sigma_{84}^4$\citep{Tremaine2002} we finally have 
\be
\Gamma_{\rm gal}(M_S) = 1.2\times10^{-10} {\rm yr}^{-1} (r_i/r_{\rm dyn})^{31/14}  \eta_{0.25}^{-23/28} \tilde{\kappa}_{2.5}^{31/14} m_{20}^{-3/14} M_{4e6}^{9/28}
\ee

Thus far, this agrees relatively well with \citet{Kocsis2012} and \citet{OLeary2009}.

\subsection{Averaging Over Many Galaxies}
From the scatter in the inferred nuclear star cluster relaxation time $T_r$ for the low $\sigma_{\rm disp}$ galaxies given in Figure 1 of \citet{Merritt2007}, \citet{OLeary2009} and \citet{Kocsis2012} claim that the variance of the central number density scales the average rate per galaxy by a factor $\xi = \overline{n^2}/\overline{n}^2 \go 30$, increasing their total rate substantially.

Here, in calculating the average rate over many galaxies, we will carefully consider the effect of variation in both the $M-\sigma$ relation and the scaling of the central density implied by \citet{Merritt2007} -- related to  the parameter $r_i/r_{\rm dyn}$ -- and show that such a substantial increase in the inferred rate is not warranted. 

{\bf Variation in the $M-\sigma$ relation:} We begin by taking the generous assumption that the intrinsic scatter in the $M-\sigma$ relation is $\sim 0.5$ dex in $M_S$ \citep{Tremaine2002}. We take the distribution of $M_S$ for fixed $\sigma_{\rm disp}$ to be log-normal such that $M_{4e6} = C_{M\sigma}\times \sigma_{84}^4$ where $C_{M\sigma}$ is a random variable with log-normal probability distribution with geometric mean $\langle C_{M\sigma} \rangle = 1$ and scale factor $\delta_{M\sigma} = \ln \sqrt{10}$. 

Our single galaxy rate (\ref{1galrate}) has scaling such that for a fixed SMBH mass bin
$\Gamma_{\rm gal}(M_S, C_{M\sigma}) \propto C_{M\sigma}^{-31/28}$. Using the properties of log-normal random variables the scaling to the average rate over this distribution is given as
\be
\xi_{M\sigma} \equiv \frac{\overline{C_{M\sigma}^{-31/28}}}{\langle C_{M\sigma}\rangle^{-31/28}} =   \exp \left[ \frac{1}{2}\left(\frac{31}{28}\right)^2\delta_{M\sigma}^2\right]   \simeq 2.25.
\ee
where $\overline{f}$ denotes averaging of $f$ over the distribution.

{\bf Variation in $r_i/r_{\rm dyn}$:} The ratio of the radius of influence, $r_i$,  to the dynamical radius, $r_{\rm dyn}$, determines the relative density of the nuclear cluster, and varies for different N-body models from $\sim 0.1 - 1$. \citep{Binney2008}. For simplicity we will take this ratio to be a log-normal distributed random variable independent of $C_{M\sigma}$.  

Figure 1 of \citet{Merritt2007} showed an estimate of the relaxation time at the dynamical radius as a function of the velocity dispersion $\sigma_{\rm disp}$ for nuclei of early-type galaxies in the ACS Virgo cluster survey \citep{Cote2004}. While the majority of the scatter in this distribution is for low-luminosity unresolved nuclear star clusters, an indication that much of this scatter may be due observational uncertainty, for the purposes of this discussion we will assume, as in \citet{OLeary2009}, that this scatter is intrinsic. We will again assume, for simplicity, that the relaxation time $T_r$ for fixed $\sigma_{\rm disp}$ is distributed log-normally, with a generous estimate of 1.5 dex standard deviation for the scatter in $\log T_r$, such that scale factor $\delta_{T_r} \simeq 1.5 \ln 10$.

 The nuclear relaxation time is given by
\be
T_r(r_{\rm dyn}) \simeq \frac{0.34 \sigma_{\rm disp}^3}{G^2 n(r_{\rm dyn}) \ln \Lambda} = \frac{0.34\sigma_{\rm disp}^3}{G^2 \ln \Lambda}\frac{4\pi r_{\rm dyn}^3}{2 \kappa M_S}
\ee
\citep{Spitzer1987}, where $\ln \Lambda$ is the Coulomb logarithm. For fixed $\sigma_{\rm disp}$ we can rewrite the $T_r$ in terms of the random variables $C_{M\sigma}$ and $r_i/r_{\rm dyn}$. 
\be
T_r(r_{\rm dyn}) \simeq  \frac{0.68\pi G M_S^2}{\ln \Lambda \kappa \sigma_{\rm disp}^3} \left(\frac{r_i}{r_{\rm dyn}}\right)^{-3} \sim \sigma_{\rm disp}^5 C_{M\sigma}^2 \left(\frac{r_i}{r_{\rm dyn}}\right)^{-3}.
\ee
This gives, for independent log-normal random variables $C_{M\sigma}$ and $r_i/r_{\rm dyn}$, 
\be
\delta_{T_r}^2 = (2\delta_{M\sigma})^2 + (3 \delta_{\rm dyn})^2
\ee
where $\delta_{\rm dyn}$ is the standard deviation of $\ln r_i/r_{\rm dyn}$. This then gives the scaling due to variation in $r_i/r_{\rm dyn}$ to the average rate of
\be
\xi_{\rm dyn}\equiv \frac{\overline{(r_i/r_{\rm dyn})^{31/14}}}{\langle r_i/r_{\rm dyn} \rangle^{31/14}} = \exp \left[\frac{1}{2} \left( \frac{31}{14}\right)^2 \delta_{\rm dyn}^2 \right]=  \exp \left[\frac{1}{2} \left(\frac{31}{14}\right)^2 \frac{\delta_{T_r}^2 - 4\delta_{M\sigma}^2}{9} \right] \simeq 6.1.
\ee

{\bf Final Rate:} The central black hole mass function can be estimated by
\be
\Phi (M_S)  \equiv \frac{dn_{gal}}{d \ln M_S} \simeq 0.0077\, {\rm Mpc}^{-3} \times \left( \frac{M_S}{M_*}\right)^{\alpha + 1} \exp\left[ - \left(\frac{M_S}{M_*}\right)^{-\beta}\right] \label{massfn}
\ee
\citep{Shankar2004} where $\alpha \simeq -1.11$, $\beta \simeq 0.5$ and $M_* \simeq 6.4 \times 10^7 M_\odot$, assuming the local Hubble constant $H_o = 70$ km s$^{-1}$ Mpc$^{-1}$. 

Integrating our mass dependence $M_{4e6}^{9/28}$ over the mass function, we can obtain the effective density 
\be
n_{\rm gal, eff} = \int_{M_{\rm S, min}}^{M_{\rm S, max}} M_{4e6}^{9/28} \Phi(M_S) \frac{d M_S}{M_S}
\ee
with which to multiply our single galaxy rate evaluated at $M_{4e6} = 1$. The mass function (\ref{massfn}) is only constructed to be valid between $10^6 M_\odot \lo M_S \lo 5\times 10^9 M_\odot$, however we expect significant contribution from smaller galaxies. The halo mass function increases for lower mass, however for dwarf galaxies the stellar mass (and therefore SMBH mass) to halo mass ratio drops significantly and the expected number density at that mass should also fall. If we integrate down to only $M_{\rm S, min} \sim 10^6 M_\odot$ this gives us an effective density $n_{\rm gal, eff} \simeq 0.043$ Mpc$^{-3}$, while extending this mass function down to a cutoff of $M_{\rm S, min} \sim 10^4 M_\odot$ yields $n_{\rm gal, eff} \simeq 0.067$ Mpc$^{-3}$. With this in mind we take the fiducial value of the effective density to be $n_{\rm gal, eff} = n_{\rm gal,5} \times 0.05$ Mpc$^{-3}$. 

Scaling to the Milky Way where \citet{OLeary2009} assume a fiducial value of $r_i/r_{\rm dyn} \simeq 0.5$, we have our final rate of eccentric captures in galactic nuclei, 
\ba
\Gamma^{(GN, EC)}_{\rm tot} &=& \frac{4}{3}\pi d^3 n_{\rm gal, eff} \overline{\Gamma_{gal}}(M_{4e6} = 1)\\
 &\simeq&  0.6 \,{\rm yr}^{-1}\left(\frac{\xi_{M\sigma}}{2.25}\right) \left( \frac{\xi_{\rm dyn}}{6.1}\right)\left(\frac{ \langle r_i/r_{\rm dyn}\rangle}{0.5}\right)^{31/14}  \left(\frac{H}{r}\right)^{-2}  \eta_{0.25}^{-23/28} \tilde{\kappa}_{2.5}^{31/14} m_{20}^{-3/14} n_{\rm gal, 5} d_{2Gpc}^3, \label{finalrate}
\ea
within $d_{2Gpc} \times 2$ Gpc, where we take 2 Gpc for the fiducial value as it is roughly the Advanced LIGO horizon distance for $10 M_\odot$ BH-BH eccentric captures. Here we've also included an additional factor $(H/r)^{-2}$ which may increase the density for nuclear clusters where significant flattening has occurred (B. Kocsis, private communication). Assuming the generous fiducial values for $\xi_{M\sigma}$ and $\xi_{\rm dyn}$, and no significant flattening, the rate for  eccentric capture of $10 M_\odot$ BH-BH encounters is 
\be
\Gamma_{\rm BH-BH}^{(GN, EC)} \simeq 0.02 \, {\rm yr}^{-1} {\rm Gpc}^{-3}.
\ee

Less generous estimates for the intrinsic scatter, $\delta_{M\sigma} \simeq 0.3 \ln 10$ \citep{Tremaine2002}, $\delta_{T_r} \simeq \ln 10$ give $\xi_{M\sigma} \simeq 1.33$ and $\xi_{\rm dyn} = 2.52$, which reduces the above rate by a factor of $\sim 4$.

\subsection{Neutron Star Rates}
We are now (finally) ready to estimate the rates for NS eccentric captures in galactic nuclei. There is  large uncertainty in the NS production rate in nuclear star clusters, mostly due to two factors. First, the IMF is unknown and could range from the standard Salpeter IMF, to an extremely top-heavy IMF \citep[e.g.][]{Bartko2010}. Second, there is great uncertainty in the effects of mass-loss for deteriming the fraction of stars with $M_{\rm ZAMS} \go 8 M_\odot$ that will become neutron stars, and the fraction that will collapse to form black holes \citep[see e.g.][for discussion]{O'Connor2011}. 

With these considerations, we will assume that between $\sim 1-10\%$ of the stars in a nuclear star cluster will become neutron stars. The low end of this range is for a standard IMF, with low mass loss and metallicity, while the high end roughly corresponds to a top-heavy IMF with more mass loss in the NS progenitors. Assuming an average stellar mass in nuclear star clusters of $\sim 0.5 M_\odot$, this gives $\kappa_{\rm NS} \sim 0.03 - 0.3$. 

Neutron stars have only had time to segregate in galaxies for which $\sigma_{\rm disp} \lo 50$ km s$^{-1}$ \citep{Miller2009}. Unfortunately these low mass galaxies also have escape velocities $\sim 2\sigma_{\rm disp}$, and thus NS kick velocities of $\sim 100$ km s$^{-1}$ will significantly reduce the retention fraction after NS formation, much like in globular clusters. So we will assume the scaling above with no significant enhancements due to segregation or flattening.

Substituting $m_{\rm tot} = 2.8 M_\odot$, $\eta = 0.25$ and $\kappa_{\rm NS} = 0.03 - 0.3$  into (\ref{finalrate}) and continuing to use the very generous assumptions above for $\xi_{M\sigma} \simeq 2.25$ and $\xi_{\rm dyn}\simeq 6.1$ we get a rate for eccentric capture of
\be
\Gamma^{(GN, EC)}_{\rm NS-NS} \simeq 0.04 - 6 \,{\rm yr}^{-1} {\rm Gpc}^{-3}.
\ee
For $10 M_\odot$ BH - $1.4 M_\odot$ NS encounters we take $m_{\rm tot} = 11.4 M_\odot$, $\eta \simeq 0.11$, $\kappa_{\rm BH} \simeq 0.025$, $\kappa_{\rm NS} \simeq 0.03 - 0.3$, and get a rate
\be
\Gamma^{(GN, EC)}_{\rm BH-NS} \simeq 0.05 - 0.6\, {\rm yr}^{-1} {\rm Gpc}^{-3}.
\ee

However, in systems where the BH's dominate the core, a cusp of more massive objects tends to flatten out the distribution of lighter objects, compared to an isothermal density profile. \citet{OLeary2009} find that the distribution can approach a power law index of 1.5 for the lighter objects in their Fokker-Planck calculation. Repeating the above calculations for $n_{\rm NS} \propto r^{-1/2}$ as a lower bound, we find 
\be
\Gamma^{(GN,EC)}_{\rm NS-NS} \simeq 0.003 - 0.3  \,{\rm yr}^{-1} {\rm Gpc}^{-3},
\ee
noting that the rate in a single galaxy is no longer dominated by the contribution due to the innermost objects.

\section{Rates for Shattering During First Passage}
Eccentric captures, described above, provide multiple close passages, however, the cross section for eccentric capture only exceeds that for shattering flares from the first passage below $v_{\rm rel} \sim 1000$ km s$^{-1}$. Near the center of nuclear star clusters -- where the density and relative velocity dispersion are the highest -- the rate for hyperbolic passages that result in shattering flares, but are not bound through GW emission can be significant. In this Appendix we calculate this rate for shattering flares during first passage.

We begin by assuming that that the speed at infinite separation is small compared to the speed at periapse $v_{\rm rel} \ll  \sqrt{2G m_{\rm tot}/r_p}$. The maximum periapse distance for shattering is then only a function of the stellar masses, and the equation of state
\be
r_{p, s} = r_{p, s}(\eta, m_{\rm tot}, {\rm EOS}),
\ee
which can be determined through setting $\Delta E_i = E_b$  as in $\S2$ of the main text above. We can then calculate the single galaxy rate as for eccentric captures, giving
\ba
\Gamma^{(GN, SF)}_{gal} &\simeq& \int_{r_{\rm min}}^{r_i} 4 \pi r^2 \frac{\tilde{N}^2}{(4\pi r_{\rm dyn}^3)^2}\left( \frac{r}{r_{\rm dyn}}\right)^{-4} \frac{2\pi G m_{\rm tot} r_{p, s}}{v_{\rm rel}} dr\\
&\simeq& 1.3 \times 10^{-10} {\rm yr}^{-1} \left(\frac{r_{p,s}}{200\,{\rm km}}\right) \left(\frac{\tilde{\kappa}}{0.03}\right)^{5/2} \left(\frac{r_i/r_{\rm dyn}}{0.5}\right)^{5/2} \eta_{0.25}^{-5/4} m_{2.8}^{-3/2} C_{M\sigma}^{5/4} M_{4e6}^{3/4}
\ea
where we've used the $M-\sigma$ relation $M_{4e6} = C_{M\sigma} \sigma_{84}^4$ and the fact that $\langle v_{\rm rel}^{-1} \rangle \simeq 1.38 \times \langle v_{\rm rel} \rangle^{-1}$ over a Maxwellian distribution. 

Again, for the same extremely generous assumptions for the intrinsic variation in $M-\sigma$ and $T_r(\sigma)$ as we did above we can evaluate,
\be
\xi^{(SF)}_{M\sigma} = \frac{\overline{C_{M\sigma}^{5/4}}}{\langle C_{M\sigma}\rangle^{5/4}} \simeq  2.82, \qquad \xi^{(SF)}_{\rm dyn} = \frac{\overline{(r_i/r_{\rm dyn})^{5/2}}}{\langle r_i/r_{\rm dyn}\rangle^{5/2}} \simeq 9.99.
\ee

To calculate the effective density we take
\be
n^{(SF)}_{\rm gal, eff} = \int_{M_{\rm S, min}}^{M_{\rm S, max}} M_{4e6}^{3/4} \Phi(M_S) \frac{d M_S}{M_S} \simeq 0.1\, {\rm Mpc}^{-3}
\ee
which gives us our rate for shattering flare encounters of
\ba
\Gamma^{(GN, SF)}_{\rm tot} 
&\simeq& 0.5 {\rm yr}^{-1}  \xi_{M\sigma} \xi_{\rm dyn} \left(\frac{r_{p,s}}{200\,{\rm km}}\right) \left(\frac{\tilde{\kappa}}{0.03}\right)^{5/2} \left(\frac{r_i/r_{\rm dyn}}{0.5}\right)^{5/2}  \left(\frac{n^{(SF)}_{\rm gal, eff}}{0.1}\right)  \eta_{0.25}^{-5/4} m_{2.8}^{-3/2}d_{2Gpc}^3.
\ea

For $\kappa_{\rm NS} \simeq 0.03 - 0.3$ we then have the range
\be
\Gamma^{(GN,SF)}_{\rm NS-NS} \simeq 0.2 - 60\, {\rm yr}^{-1} {\rm Gpc}^{-3}
\ee
for the fiducial values above, including our optimistic estimates for $\xi_{M\sigma} \simeq 2.82$ and $\xi_{\rm dyn} \simeq 9.99$. 

Similar to the case for eccentric captures, less generous estimates for the intrinsic scatter significantly reduce these rates. Taking $\delta_{M\sigma} \simeq 0.3 \ln 10$ and $\delta_{T_r} \simeq \ln 10$, we have $\xi_{M\sigma} \simeq 1.45$ and $\xi_{\rm dyn} \simeq 3.24$, which reduces the above rates by a factor of $\sim 6$. 

If we repeat the above calculations for the flattened distribution $n_{\rm NS} \propto r^{-1/2}$ we obtain a reduced rate estimate of 
\be
\Gamma^{(GN,EC)}_{\rm NS-NS} \simeq 0.005 - 0.5  \,{\rm yr}^{-1} {\rm Gpc}^{-3}.
\ee

\renewcommand{\bibsection}{\section{References}} 
\bibliographystyle{apj}

\end{document}